\documentstyle[12pt,prl,aps]{revtex} 
\large
\def\addcontentsline#1#2#3{\relax}

\long\outer\def\demo#1. #2\par{\medbreak\noindent {\bf#1.\enspace}
        {\rm#2}\par\ifdim\lastskip<\medskipamount\removelastskip
        \penalty55\medskip\fi}
\newcommand{\ben}{\begin{enumerate}}
\newcommand{\een}{\end{enumerate}}
\def\bdemo#1. #2\par{\medbreak\noindent {\bf#1.\enspace}{\rm#2}\par}
\def\edemo{\ifdim\lastskip<\medskipamount\removelastskip\penalty55\medskip\fi}

\def\p{{\partial}}
\def\lh{\ell_H}
\def\H{{\cal H}}
\def\0{{\bf 0}}
\def\1{{\bf 1}}

\def\bfa{\mathop{a\kern-.5em{a}\kern-.5em{a}}\nolimits}
\def\bfb{\mathop{b\kern-.47em{b}\kern-.47em{b}}\nolimits}
\def\bfx{\mathop{x\kern-.6em{x}\kern-.6em{x}}\nolimits}
\begin{document}
\baselineskip 18pt
\author{Yshai Avishai$^{\,1,2}$, Yehuda Band$^{\,3}$, 
and David Brown$^{\,4}$}
\address{${}^{1}$Department of Physics, Ben Gurion University 
of the Negev,Beer-Sheva 84 105, Israel \\
${}^{2}$NTT Basic Research Laboratories, 3-1 Morinosato, Wakamiya,
Atsugi-shi Kanagawa-ken, Japan \\
${}^{3}$Department of Chemistry, Ben Gurion University of the Negev,
Beer-Sheva 84 105, Israel \\
${}^{4}$Goddard Institute for Space Studies, 2880 
Broadway, New York, NY 10025  }

\title{Conductance Distribution Between Hall Plateaus}

\maketitle

\begin{abstract}
Mesoscopic fluctuations of two-port conductance and four-port 
resistance between Hall plateaus are studied within a realistic model 
for a two-dimensional electron gas in a perpendicular magnetic field 
and a smooth disordered potential.  The two-port conductance 
distribution $P(g)$ is concave between $g=0$ and $g=1$ and is nearly 
flat between $g=0.2$ and $g=0.8$.  These characteristics are 
consistent with recent observations.  The distribution is found to be 
sharply peaked near the end-points $g=0$ and $g=1$.  The distribution 
functions for the three independent resistances in a four-port Hall 
bar geometry are, on the other hand, characterized by a central peak 
and a relatively large width.
\end{abstract}


\section {Introduction}

The dimensionless conductance $g$ of small metallic disordered systems 
at very low temperature (where the phase-coherence length is larger 
than the linear size of the system), is characterized by its 
distribution $P(g)$ which is nearly Gaussian with mean value 
proportional to the size of the system and variance of order unity.  
These properties, also termed as universal conductance fluctuations, 
are valid deep inside the metallic regime, for which $g \gg 1$.  The 
question whether such (or different) fluctuations occur also in the 
critical regime where $g$ is of the order of unity has been addressed 
in connection with the Anderson metal insulator transition prevailing 
in three-dimensional disordered systems \cite{Boris}.  In this case 
the critical region can be approached from the metallic side using 
scaling arguments and expansion in $1/g$.  Unfortunately, this is not 
possible in the quantum Hall regime, since insulating phases exist on 
both sides of the critical point.

In a recent experiment \cite{Cobden} the conductance distribution 
function at the quantum Hall transition point was elucidated.  The 
geometry of the experiment was that of a two-dimensional strip with 
two-ports, namely, one source and one drain.  The transition point is 
defined as that value of the magnetic field for which the averaged 
conductance is $(n+1/2) \frac{e^{2}} {h}$.  (Since the focus of study 
is mesoscopic fluctuations, the notion of transition point is more 
appropriate than that of critical point, which, strictly speaking, 
applies only in the thermodynamic limit).  It was found that in the 
range of values $0.1 < g < 0.9$ the distribution $P(g)$ is roughly 
flat.  Subsequent theoretical work was mostly based on the network 
model \cite{Chalker}.  The qualitative features of the experimental 
distribution have been obtained by numerical calculations \cite{Cho} 
based on the network model with free boundary conditions on the 
transverse coordinate.  Calculations were performed also with periodic 
boundary conditions in order to elucidate the role of edge states.  
Roughly speaking, adopting free (periodic) boundary conditions 
corresponds to calculating $R_{xy}$ ($R_{xx}$).  Evaluation of $P(g)$ 
using real space renormalization group methods applied to the network 
model resulted in a concave shape with a very flat bottom 
\cite{Raikh,Arovas} and sharp peaks near $g=0$ and $g=1$.  It is 
therefore interesting to check (against other models) whether the 
network model which proves to be extremely successful in describing 
the critical behavior of the integer quantum Hall effect is also 
capable of encoding the correct statistical properties of the 
conductance for mesoscopic systems.

A natural extension of the physics discussed above is related to the 
mesoscopic fluctuations of the Hall resistance.  To the best of our 
knowledge, investigation of the Hall resistance fluctuations in the 
transition points between Hall plateaus has not yet been  
reported.  It might be worth pointing out here that the precise 
relation between the two-port conductance $g$, discussed in connection 
with the experiment of Ref.~\cite{Cobden}, and the conductivity tensor 
$\sigma_{\mu \nu}$, defined through the Kubo formula, is not 
unambiguously clear.  In particular, an interesting sum-rule relates 
the two-port transmission coefficient to a {\em combination} of Hall 
and dissipative resistivities \cite{Streda}.  This relation holds, in 
fact, not only at the transition region between plateaus.  In order to 
separate the effects of the Hall and dissipative resistivities a {\em 
multi-probe} measurement is required, as employed in the original 
experiment \cite{Klitz}.  Once a multi-probe geometry is defined, 
numerous resistances can be determined (beside the Hall resistance).

In order to address these questions, calculations of conductance 
(resistance) distributions between Hall plateaus for two (four) port 
setups are presented below (in the four-probe geometry, the 
appropriate quantity is in fact the resistance).  The Schr\"{o}dinger 
equation for an electron moving in part of the plane restricted by 
certain boundaries and subject to a strong perpendicular magnetic 
field and a smooth disordered potential is solved, and transmission 
and reflection coefficients between various ports are computed 
\cite{Schult,AB}.  The two-port conductance is calculated using the 
Landauer \cite{Landauer} formula, while the various four-probe 
resistances are calculated from the B\"{u}ttiker formula 
\cite{Markus}.  Note that no projection on lowest Landau level is 
made.

We find that the distribution $P(g)$ of the two-port conductance is 
concave, with rather shallow minimum at $g=1/2$.  As we have already 
pointed out, the experimental data indicate a flat distribution 
between $g=0.1$ and $g=0.9$, and in this domain the calculated 
distribution is quite smooth and hence qualitatively accounts for the 
experimental distribution.  On the other hand, there is so far no 
experimental evidence for the occurrence of two peaks near $g=0$ and 
$g=1$.  It is also found that conductances $g(E_{F},H)$ at different 
points of the fermi energy ($E$) and magnetic field ($H$) are 
correlated if the two points are along a transition line.

In the four-probe geometry there are three independent resistances 
that can be measured, $R_{12,34}$ $R_{13,24}$ and $R_{13,13}$.  In 
this notation $R_{ij,kl}$ is the ratio of the voltage difference 
between ports $i$ and $j$ and the current flowing between ports $k$ 
and $l$.  Inspecting the histograms of these three independent 
resistances at the transition point (defined by $<R_{13,24}>=0.6 
\frac{h} {e^{2}}$), show that all three of them have a distribution 
which is, more or less, close to a normal one.  The width of 
$R_{12,34}$ (the so called bend resistance) is of order $\Delta R = 
0.2 h/e^{2}$ while for $R_{13,24}$ (the Hall resistance), and 
$R_{13,13}$ the order of the width is $\Delta R = 0.5 h/e^{2}$.  This 
is quite large compared with the corresponding average.  Thus, within 
the realm of mesoscopic fluctuations, the Hall resistance is not a 
universal number.  We are unaware of an experiment aimed at studying 
mesoscopic fluctuations of multi-probe resistances.

The method of calculation is briefly explained in Sec.~$II$, and 
the results are presented and discussed in Sec.~$III$.

\section{Formalism}

The two-port system is schematically displayed in Fig.~\ref{fig1}a.  
It consists of a strip $-\infty < x < \infty, -L/2 \le y \le L/2$ 
which confines the motion of electrons whose wave function vanishes on 
the walls.  The four-port system is schematically displayed in 
Fig.~\ref{fig1}b.  It consists of two perpendicular strips of width 
$L$ which intersect each other.  It is convenient to number the ports 
(leads) from the left in a counter clockwise sense, so lead number $1$ 
is the left lead {\em etc.}.  In both two and four-port systems, the 
motion of the electron is ballistic except within the domain $(-L/2 
\le x \le L/2, -L/2 \le y \le L/2)$ where the electron is subject to a 
smooth disordered potential, defined below.  The system is subject to 
a strong perpendicular magnetic field.  It is useful in this geometry 
to define the vector potential ${\bf A}$ in the Landau gauge, $A_x = 
-Hy$, where $H$ is the magnetic field strength On this square one 
considers a lattice of $N^2$ Gaussian potentials of width $\sigma$, 
centered at the points 
$(x_n,y_m) = \big(-L/2+a/2+(n-1)a,-L/2+a/2+(m-1)a\big)$, 
with $n,m=-N/2+1,\cdots,N/2$, assuming that $N=L/a$ is an even 
integer.  The random potential is written as
\begin{eqnarray}
V(x,y)=\sum_{nm} v_{nm}
\exp\left(-\frac{(x-x_n)^2+(y-y_m)^2}{2\sigma^2}\right).
\label{eq_pot}
\end{eqnarray}
The impurity strengths $v_{nm}$ are independent random energies, 
distributed (for convenience) uniformly between $-w/2$ and $w/2$.  For 
this distribution one has
\begin{eqnarray}
\langle v_{nm}\rangle=0 ,
\label{eq_vav}
\end{eqnarray}
and
\begin{eqnarray}
 \langle v_{nm}v_{n'm'}\rangle
=\frac{w^2}{12}\delta_{nn'}\delta_{mm'}.
\label{eq_vcorr}
\end{eqnarray}
The single-particle reduced Hamiltonian then reads
\begin{eqnarray}
\H=\frac{\hbar^2}{2m} \left [\left(-i\frac{\p}{\p x}-
\frac{y}{\lh^2}\right)^2
-\frac{\p^2}{\p y^2}\right ] + V(x,y)
\equiv \H_{0} + V .
\label{eq_Ham}
\end{eqnarray}
The magnetic length is given by $\lh=(\hbar c/eH)^{1/2}$, where 
$-e<0$ is the charge of the electron.

In order to calculate transmission and reflection coefficients one 
needs to obtain the asymptotic states away from the central region, 
i.e., in the leads.  Consider for example the left lead $-\infty < x < 
-L/2, -L/2 < y < L/2$.  For a given Fermi-energy $E_{F}$ the solution 
of the free Schr\"{o}dinger equation $\H_{0} \Psi=E_{F}\Psi$ in this 
domain is given by
\begin{eqnarray}
\psi_{n}^{\pm} (x,y)=q_{n}^{-1/2}e^{\pm i q_{n} (x+L/2)}f(y;\pm q_{n}) ,
\label{eq_psi}
\end{eqnarray}
where $f(y;q)$ is a linear combination of parabolic cylinder functions 
\cite{Abram} such that the boundary conditions $f(-L/2;q)=f(L/2;q)=0$ 
determine the eigen-momenta $q_{n}(E_{F},H)$.  Note that in the case 
of a four-port system, the role of $x$ and $y$ is interchanged in the 
vertical leads.  For convenience, the amplitude of the incident plane 
wave is set to be exactly unity on the boundaries between the ideal 
conductors and the region of impurities.

It is worth mentioning that for the present geometry, it is not 
possible to impose periodic boundary conditions in the transverse 
direction.  This situation is distinct from the one encountered in the 
network model \cite{Chalker}.

The number $M$ of open channels is the maximum number of positive 
momenta.  It is an increasing number of the Fermi-energy $E_{F}$ and a 
decreasing number of the magnetic field $H$.  The behavior of the 
momenta $q_{n}(E_{F},H)$ as function of the two parameters has been 
discussed elsewhere \cite{Schult}.

Consider now a solution of the full scattering problem such that there 
is an incoming wave $\psi_{n}^{+}$ of unit amplitude in one of the 
leads (say lead $i$), and outgoing waves in all leads $j$ (including 
the lead $j=i$).  The wave function in lead $j$ is then given by,
\begin{eqnarray}
\Psi_{n}^{ji}=\sum_{m=1}^{M} \psi_{m}^{+}t_{mn}^{ji}, (for j \ne i) ,
\label{eq_Psiji}
\end{eqnarray}
and 
\begin{eqnarray}
\Psi_{n}^{ii}=\psi_{n}^{+}+\sum_{m=1}^{M} \psi_{m}^{-}t_{mn}^{ii}.
\label{eq_Psiii}
\end{eqnarray}
The matrices $t_{mn}^{ji}$ and $t_{mn}^{ii}$ are transmission 
reflection amplitudes respectively.  Within linear response formalism, 
the two-port conductance $g$ and the four-port resistance tensor 
$R_{ij,kl}$ can be determined by these amplitudes.  In the first case 
the relation is given by $g=Tr[t^{12}t^{12\dag}]$ while in the second 
case the relation is given in Ref.~\cite{Markus}.  The essential part 
is: $R_{ij,kl} \propto (T^{ki}T^{lj}-T^{kj}T^{li})$, where 
$T^{ki}=Tr[t^{ki}t^{ki\dag}]$.

We recently introduced a variational algorithm for the generalized Z-matrix
which, given a finite region in configuration space, effects the following
transformation of the scattering wave-function on the boundary surface:
\begin{equation}
{\bf Z}(a\Psi(S) + bD\Psi(S)) = d\Psi(S) + qD\Psi(s).
\label{eq:zmat}
\end{equation}
In the above we define the generalized normal derivative 
$D\Psi(S) = {\bf n}\cdot{{i}\over{\hbar}}{\bf\pi}\Psi(S)$, where
${\bf\pi} = {\bf p}-{{e}\over{c}}{\bf A}$ is the generalized momentum
for a particle of momentum ${\bf p}$ and charge $e$ in a vector
potential ${\bf A}$. The parameters $a$, $b$, $d$, and $q$ are 
arbitrary except that they obey the constraint $aq-bd=-1$. In all studies
we use $a=-q=1$, $b=d=0$, for which ${\bf Z}$ is the negative of the
log-derivative matrix. Actually, we use the inverse of this matrix, as
discussed in reference \cite{Brown}, which is the negative of the well-
known ${\bf R}$ matrix. The notation $\Psi(S)$ indicates that the 
wave-function is evaluated on the boundary surface $S$ and projected onto 
a suitable basis defined on the surface. In all studies we use sine
functions on the boundaries of the regions in Fig.~1. Using the coefficients
of $f(y;\pm q_n)$ in the sine basis,\cite{Tamura} we can plug Eqs.~(6) 
and (7) evaluated on the boundaries of the scattering structures into 
Equation (\ref{eq:zmat}) and solve for the coefficients $t^{ij}_{mn}$ 
using standard singular-value decomposition algorithms \cite{Pressetal}.

Because of the large bases required, direct computation of the 
R-matrix was not convenient.  Instead, we used a straightforward 
extension of a recursive cellular R-matrix algorithm outlined by 
Nesbet \cite{Nesbet}.  In the 2-port case, we divided the coordinate 
$x$ into $N_x$ subregions.  On each subregion, we computed an R-matrix 
using a basis consisting of sine functions in $y$ which go to zero at 
$y=\pm{{l}\over{2}}$ and a Legendre discrete variable representation 
(DVR) in $x$ \cite{Light} (for a discussion of the Legendre DVR we 
used, see Ref.~\cite{Brown}).  Using the continuity and 
differentiability requirements on $\Psi$ at the boundaries between 
cells, we concatenated adjacent R-matrices into global R-matrices for 
the entire region, as discussed by Nesbet.

In the 4-port case, we divided the region in Fig.~1b into $N^2_s$ 
square subregions; that is, we divided $x$ and $y$ each into $N_s$ 
subregions.  On each subregion we computed an R-matrix as discussed in 
Ref.~\cite{Brown} and concatenated them into the global R-matrix for 
the entire region in a procedure analogous to that outlined above.  A 
few cautions may be in order here.  First, for the subregion 
R-matrices we use Legendre functions as surface functions, not sine 
functions.  This modification is necessary since to use sine functions 
on the subregions would be to imply that the scattering wave-functions 
are zero at all the points of intersection between any four cells.  
Second, the scattering wave-function is zero at the four corners of 
the square in Fig.~1b.  These conditions are enforced by means of 
linear constraints on the wave-functions in the subcells at these 
corners.  Finally, since the final global R-matrix computed in the 
manner outlined above would be expressed in terms of Legendre 
functions confined to $4N_s$ subregions on the surface, the resulting 
matrix must be projected onto the desired surface sine functions.

\section{Results}

Let us now make a choice of the potential parameters.  We first assign 
a relation between $a$ and $\sigma$ such that the exponential factor 
in the overlap integral between Gaussians of two adjacent impurities, 
calculated to be $2\pi\sigma^2\exp\left(-\frac{a^2}{4\sigma^2}\right)$ 
equals $\exp(-1/2)$.  Hence, $\sigma=a/\sqrt{2}$.  Next, the length of 
the system $L$ and the density of impurities $\rho \equiv (Na/L)^{2}$ 
are determined by the requirement that in the range of pertinent 
magnetic fields, the length of the system is much larger than the 
magnetic length and there is about one impurity per magnetic area 
$\lh^2$.  We have found that computation time grows quickly with the 
total number of impurities, $N^2$.  The present calculations are 
carried out with $N=10$.  This yields $L \approx 10 \lh$.

Hereafter we use atomic units; length is given in units of the Bohr 
radius $r_{0}=0.5292$ Angstrom, energy is expressed in Hartree (1 
Hartree $(e^2/r_{0})\,=\,27.2$ eV), and the atomic unit of magnetic 
field is equivalent to $1715.1$ Tesla.  Calculations are performed for 
system width $L=2078$ Bohr.  The value of disorder strength ($w$) must 
be commensurate with those of Fermi-energy and magnetic field so that 
the conductance is mainly dominated by magnetic field effects and not 
by disorder.  We chose $w=4.26 \times 10^{-5}$ Hartree and require 
that the Fermi-energy $E_{F}$ and the magnetic field $H$ will vary in 
the region for which the transition between the first plateaus to the 
insulating phase occur.  These are $1.3 \times 10^{-5}< E_{F} < 1.5 
\times 10^{-5}$ and $3.45 \times 10^{-3} < B < 3.80 \times 10^{-3}$ 
respectively.

\subsection{Two-Port System}

First we show in Fig.~\ref{fig2} the two-port average conductance 
together with its variance near the last plateau.  This figure shows 
that mesoscopic fluctuations occur mainly in the region of transition 
between plateaus, in agreement with experiment \cite{Cobden}.  The 
distribution $P(g)$ of the two-port conductance is displayed in 
Fig.~\ref{fig3} for three values of the magnetic field for which the 
average value of the conductance is $0.514$, $0.498$ and $0.478$.  In 
all three cases the distribution is concave with very shallow minimum 
at the center and maxima at the endpoints $g=0$ and $g=1$.  The region 
between $g=0.2$ and $g=0.85$ is roughly flat, in agreement with 
experiment \cite{Cobden}.  Note that the distribution presented here 
is similar to the one predicted from the network model using the 
renormalization group technique \cite{Raikh,Arovas}.  On the other 
hand, its characteristics near the end-points is distinct from the one 
obtained by numerical solution of the network model (see Fig.~5b in 
Ref.~\cite{Cho}).  Assuming the latter procedure to be more accurate 
than the renormalization group one, we suspect that the network model 
and the present strip geometry model yield distributions which are 
very similar in the main part of the interval $0<g<1$, but are 
disparate around the end-points.  On the other hand, the results of 
the present model are in good agreement with RG analysis of the 
network model.  The question which one is closer to the real 
distribution requires further experimental study.

Another question which is interesting in this context is whether 
conductances $g(E_{F},H)$ at different points in the $(E_{F},H)$ plane 
are correlated.  Preliminary results indicate that the answer to this 
question is affirmative \cite{Cobden1}.  A useful concept in 
mesoscopic physics is that of ergodicity, which roughly speaking, 
implies that fluctuations (of various quantities such as the 
conductance) as function of a given parameter are equivalent to 
sample-to-sample fluctuations.  The experiment discussed here 
\cite{Cobden} is based on this hypothesis, since measurements are 
performed on a single sample at many values of the magnetic field.  In 
other words, it is tacitly assumed that, in the thermodynamic limit, 
$\langle g(E_{F},H) g(E_{F}',H')$ decreases very rapidly with 
increasing $|E_{F}-E_{F}'|$.  The question that can be posed here is 
whether nontrivial correlations $\langle g(E_{F},H) g(E_{F}',H') 
\rangle$ do exist.  Within the limitation on system size as explained 
above it is not possible to give a definite answer here.  Instead, we 
display in Fig.~\ref{fig4} a contour plot of $\partial g / \partial 
H$ in the $(E_{F},H)$ plane.  The regions of maximal derivatives are 
clearly shown in the projected contour plot.  In the bulk geometry 
they simply correspond to the Landau levels $E_{n}=(n+ \frac {1} {2}) 
\hbar \omega$.  Here they correspond to transition between $n$ and 
$n+1$ edge states.  It might then be argued that if $(E_{1},H_{1})$ 
and $(E_{2},H_{2})$ are two different points along one of these lines 
then the correlation $\langle g(E_{1},H_{1}) g(E_{2},H_{2}) \rangle$ 
is larger than otherwise.  Yet, this analysis is only qualitative.  In 
particular, the experimental analysis of conductance derivative 
reveals correlation lines which cross each other \cite{Cobden1}.  
Interpretation of these latter observations apparently goes beyond the 
single particle picture.

\subsection{Four-Port System}

We now present results for resistance distributions in the four-port 
geometry.  As mentioned in the introduction, there are three 
independent resistances to be studied, $R_{13,24}$ (the Hall 
resistance), $R_{12,34}$ (the bend resistance) and $R_{13,13}$.  Note 
that the latter should not be confused with the dissipative 
conductance $R_{xx}$ since current and voltage are measured at the 
same points in the leads.  Calculations (and measurement) of $R_{xx}$ 
require a six port geometry.  This is the reason why, in the present 
scheme, we can calculate the Hall resistance and not Hall conductance.  
Nevertheless, the resistance $R_{13,13}$ has a non-trivial physical 
content of its own \cite{AB}.

The value of the magnetic field $H_{c}$ at which the quantum Hall 
transition occurs is usually determined by that value for which 
$R_{xx}$ has a sharp peak, or, equivalently, the Hall conductance 
takes on the values $G_{xy} = n+1/2$.  The corresponding value of the 
Hall resistance $R_{xy}=\frac {G_{xy}}{G_{xx}^{2}+G_{xy}^{2}}$ at the 
transition point depends on the corresponding value of $G_{xx}$.  
Assuming $G_{xx}(H_{c})=0.5$ \cite{Huo}, implies that at the 
transition point between $n=2$ (where $R_{xy}= \frac {h} {2 e^2}$) and 
$n=1$ (where $R_{xy}=\frac {h} {e^2}$) one has $R_{xy}(H_c)=0.6 \frac 
{h} {e^2}$ (and not $0.75 \frac {h} {e^2}$ as one might be tempted to 
think).  We choose to present our results at this point and not the last 
transition point since there $R_{xy}=\frac {h} {2 e^2}$ exactly as in 
the plateau.

In Fig.~\ref{fig5} we display the distributions of the three 
resistances at $H_{c}$.  All the resistances are characterized by a 
central peak and small tails.  In all three cases, the width of the 
distribution is of the same order as the average itself, which means 
that mesoscopic fluctuations are rather significant.  We are tempted 
to conjecture that the distribution of the Hall conductance would be 
similar to that of the Hall resistance, which, comparing with 
Fig.~\ref{fig3} is much distinct from that found for a two-port 
system.  This is quite significant since, based on the edge state 
picture, it is intuitively suggestive that two-port 
magneto-conductance at strong magnetic field is the same as the Hall 
conductance, which just counts the number of edge states.  This is 
definitely true in the plateau regions, and apparently also in the 
transition region if the {\em average} quantity is considered.  On the 
other hand, the pertinent mesoscopic fluctuations are distinct.  We 
have yet to find an explanation for this difference.  Unfortunately, 
measurements of resistance fluctuations in multi-port systems are not 
yet available.

In conclusion, we have presented a study of conductance and resistance 
distribution functions in the quantum Hall regime, using a realistic 
calculation scheme.  The two-port conductance distribution agrees well 
with the experimental one \cite{Cobden}.  Its behavior is 
characterized by a concave shaped curve which is rather flat at the 
center and sharply peaked near the end-points $g=0$ and $g=1$.  Our 
result also indicate possible correlations of conductances at 
different points in the $(E,H)$ plane, if these points are located on 
a transition line.  The four-port resistance distributions are peaked 
at their respective central values and have large widths (compared 
with their mean values).  It remains to be determined whether these 
fluctuations survive in the thermodynamic limit at zero temperature.

\begin{acknowledgments}
This work was supported in part by grants from the Israel Academy of 
Science {\em Centers of Excellence}, the DIP program for German Israel 
Scientific Cooperation, and the US-Israel Binational Science 
Foundation.  We are grateful to J. T. Chalker D. Cobden, E. Kogan and 
D. Shahar for discussions and suggestions.
\end{acknowledgments}

\section{Figure Captions}

\begin{figure}
\caption{Sample geometry used in the present calculations.  a) Two 
port system and b) four-port system.  The shaded area contains a 
disorder potential.
\label{fig1}
}
\end{figure}

\begin{figure}
\caption{two-port average conductance (solid line) and its variance 
(dash-dotted line) as a function of the magnetic field for fixed Fermi 
energy $E=3.65 \times 10^{-3}$ Hartree.  The system width is $L=2078$ 
Bohr, and the disorder strength is $w=2.3 \times 10^{-5}$ Hartree.
\label{fig2}
}
\end{figure}

\begin{figure}
\caption{Distribution $P(g)$ of two-port conductance in the middle of 
the transition between Hall plateaus $n=1$ and $n=0$ for three 
different values of the magnetic field, $B_{1}=0.00360$ a.u., 
$B_{2}=0.00365$ a.u.  and $B_{3}=0.00370$ a.u.  The corresponding 
values of the average conductances are $g_{1}=0.514$, $g_{2}=0.498$ 
and $g_{3}=0.478$.  Parameters are as in figure \ref{fig2}.
\label{fig3}
}
\end{figure}

\begin{figure}
\caption{Three dimensional and contour plot of the conductance 
derivative $\partial g(E_{F},H)/ \partial H$ in the $(E_{F},H)$ plane 
for a given sample.  Length and disorder parameters are as in 
Fig.~\ref{fig2}. Units of energy and magnetic fields are 
specified in the text, while $g$ is in units of $e^2/h$. 
\label{fig4}
}
\end{figure}

\begin{figure}
\caption{Displayed from top to bottom are distributions of Hall resistance 
$R_{13,24}$, bend resistance $R_{12,34}$ and the ``through'' 
resistance $R_{13,13}$ for $H=0.00135$ a.u. for which the average value 
of the Hall reistance is $0.6$. 
\label{fig5}
}
\end{figure}

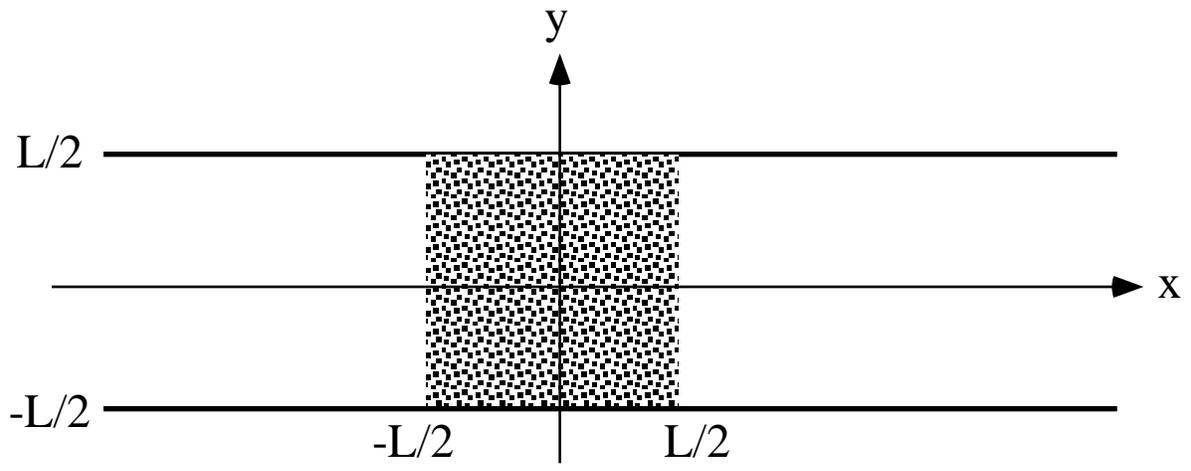
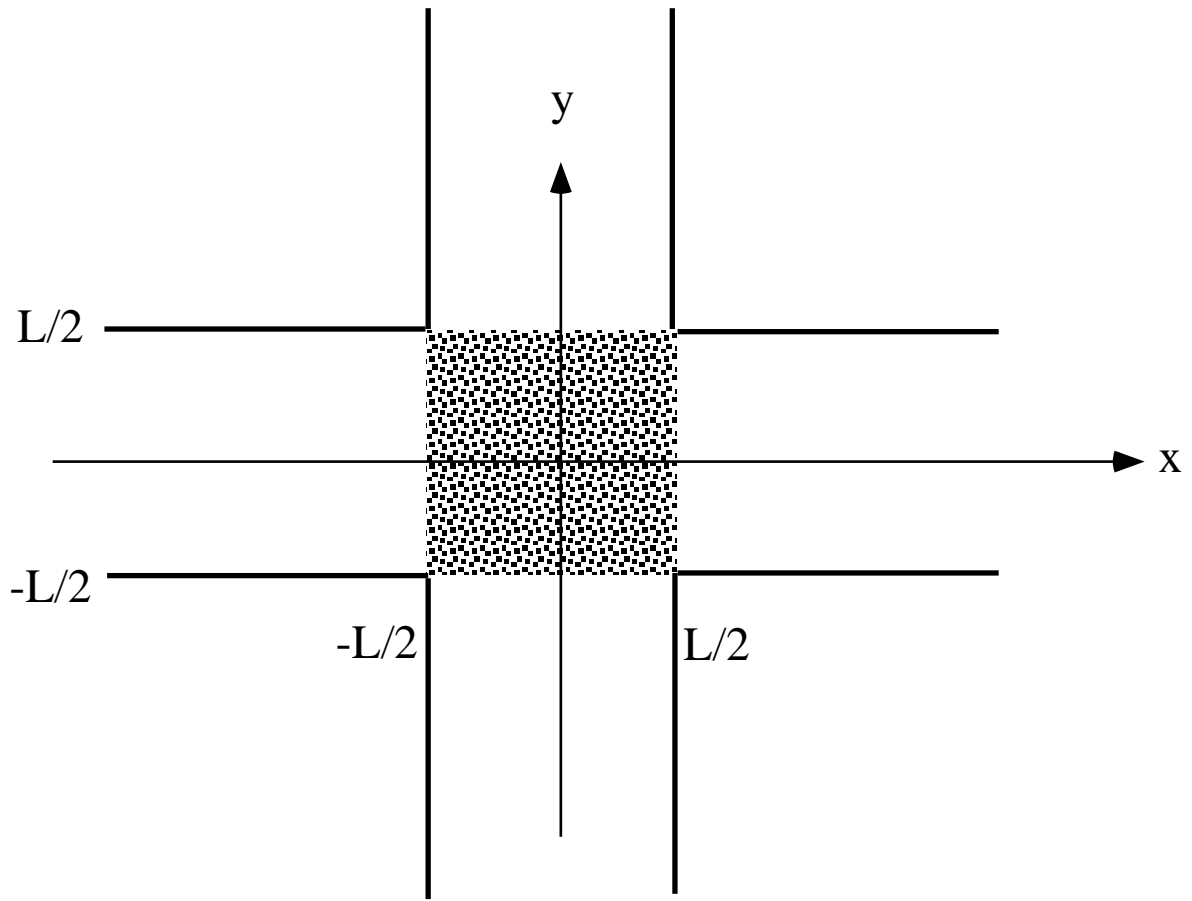
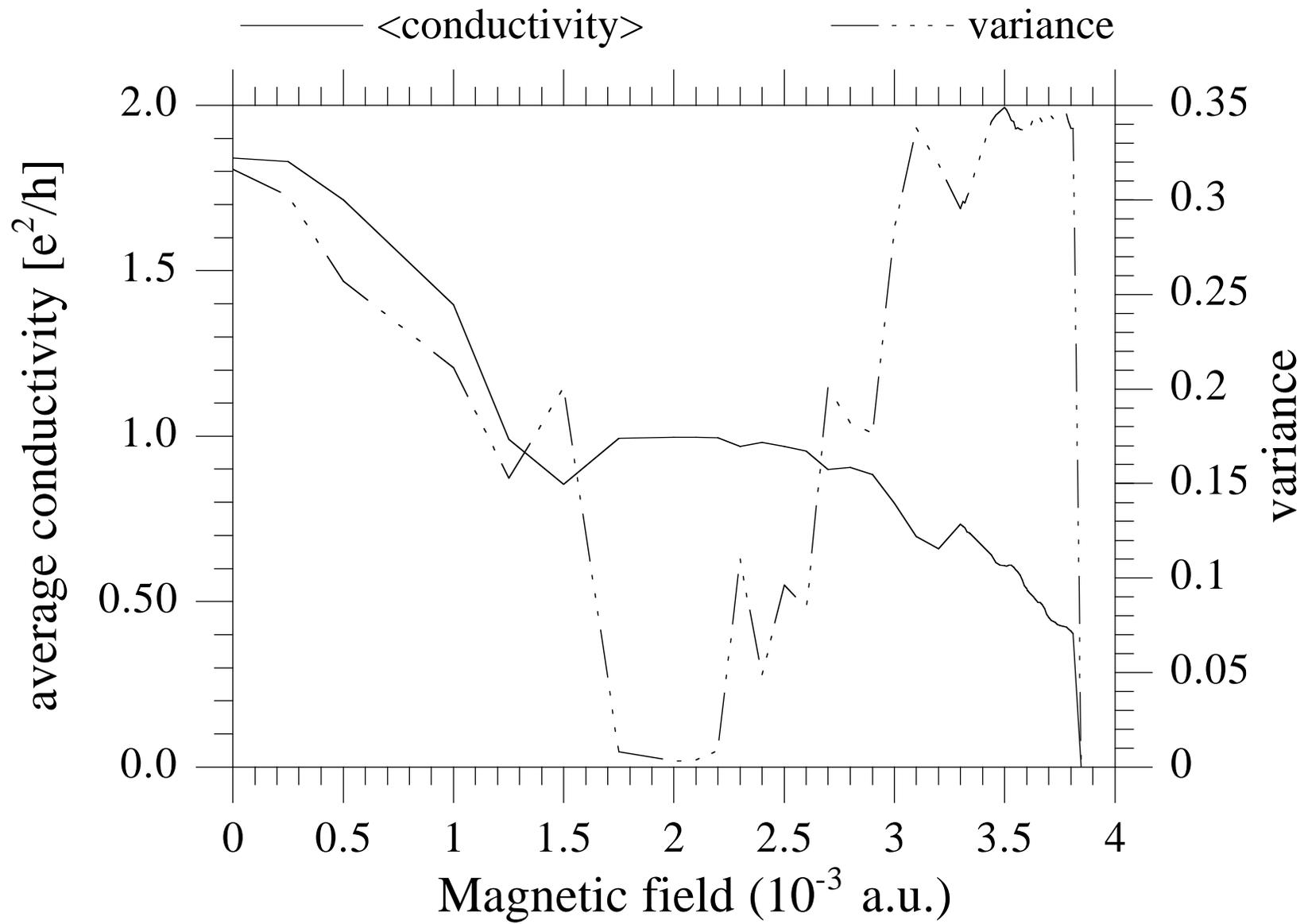
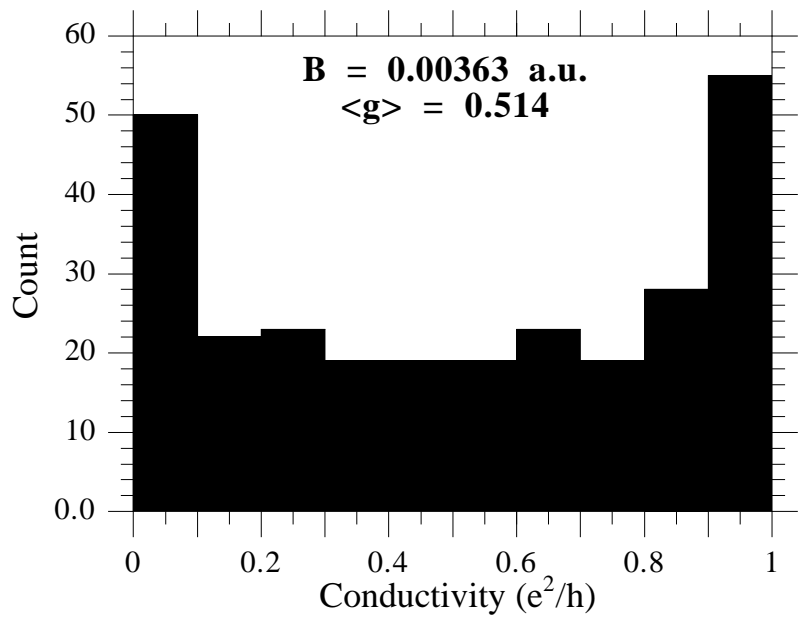
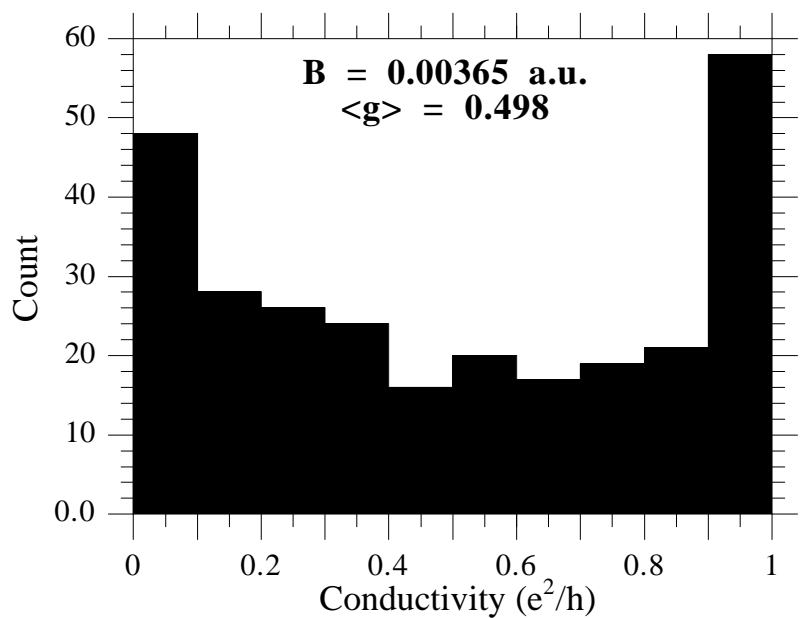
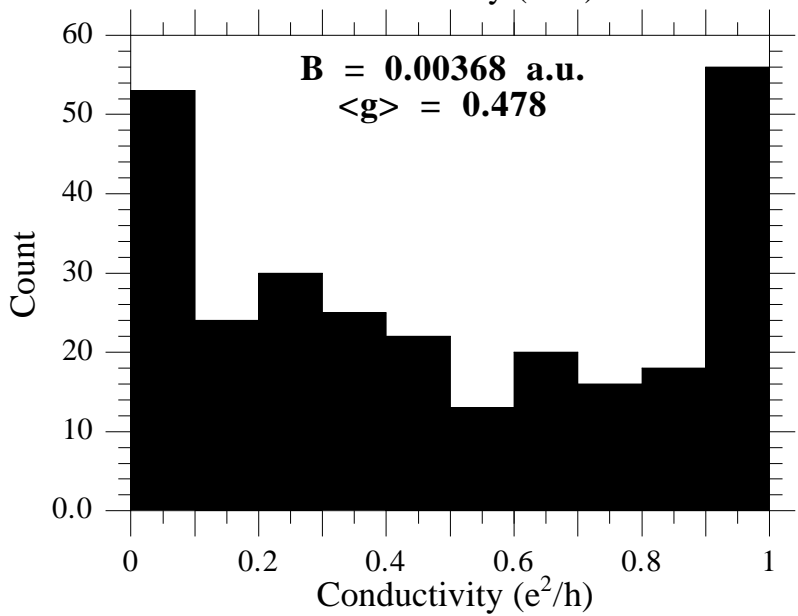
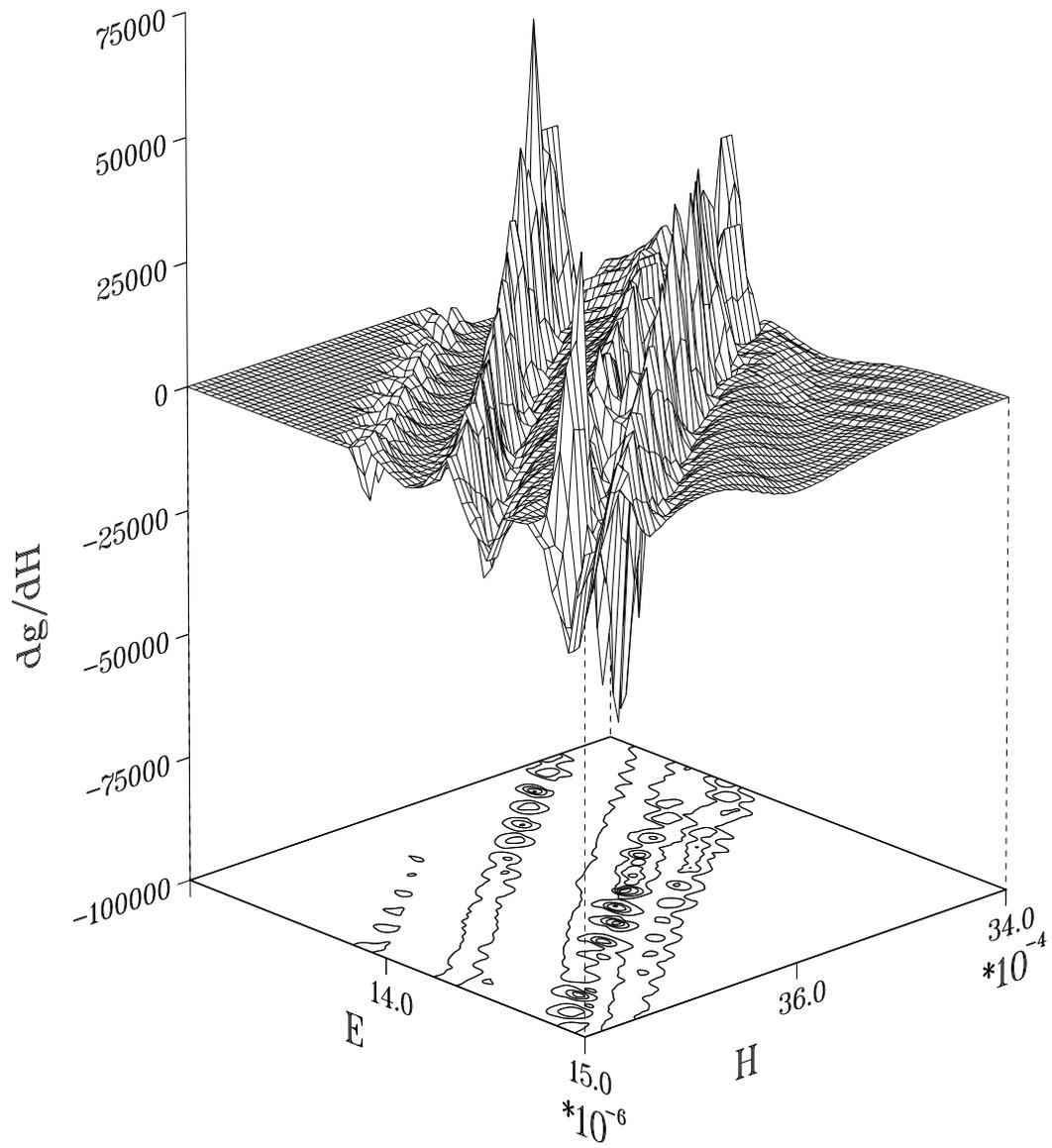
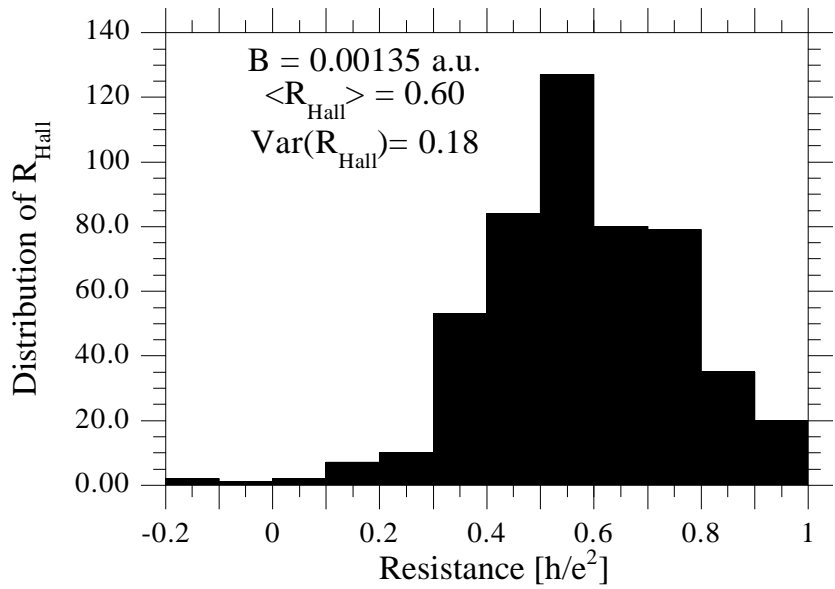
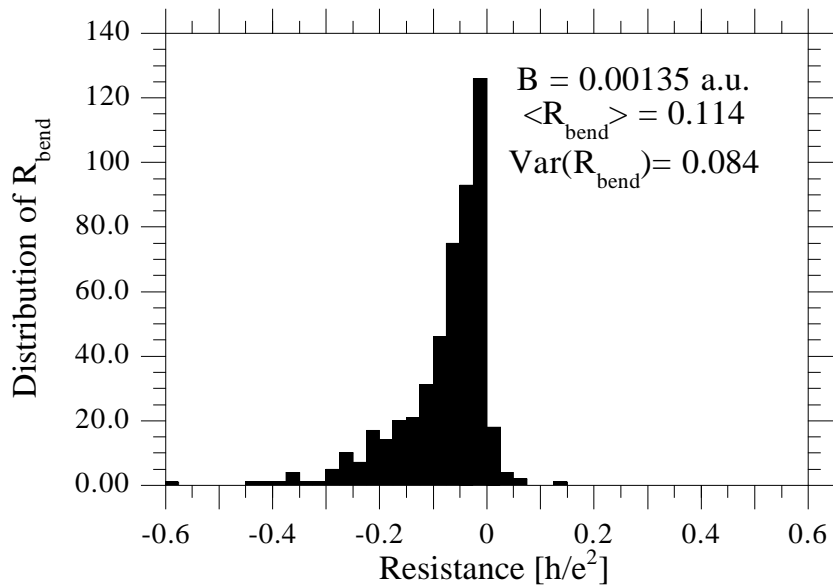
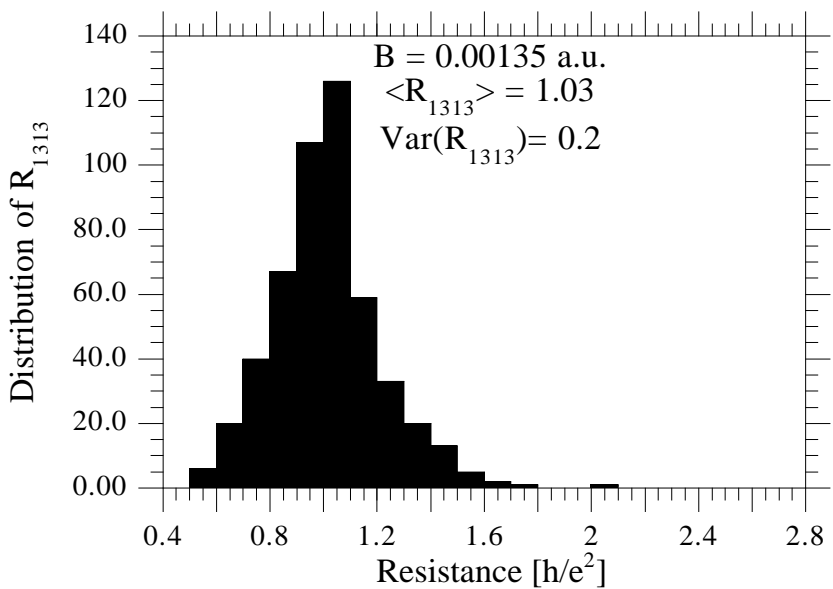
\end{document}